\PassOptionsToPackage{unicode}{hyperref}
\PassOptionsToPackage{hyphens}{url}
\PassOptionsToPackage{dvipsnames,svgnames,x11names}{xcolor}
\documentclass[
]{article}
\usepackage{amsmath,amssymb}
\usepackage{lmodern}
\usepackage{iftex}
\ifPDFTeX
  \usepackage[T1]{fontenc}
  \usepackage[utf8]{inputenc}
  \usepackage{textcomp} % provide euro and other symbols
\else % if luatex or xetex
  \usepackage{unicode-math}
  \defaultfontfeatures{Scale=MatchLowercase}
  \defaultfontfeatures[\rmfamily]{Ligatures=TeX,Scale=1}
\fi
\IfFileExists{upquote.sty}{\usepackage{upquote}}{}
\IfFileExists{microtype.sty}{% use microtype if available
  \usepackage[]{microtype}
  \UseMicrotypeSet[protrusion]{basicmath} % disable protrusion for tt fonts
}{}
\makeatletter
\@ifundefined{KOMAClassName}{% if non-KOMA class
  \IfFileExists{parskip.sty}{%
    \usepackage{parskip}
  }{% else
    \setlength{\parindent}{0pt}
    \setlength{\parskip}{6pt plus 2pt minus 1pt}}
}{% if KOMA class
  \KOMAoptions{parskip=half}}
\makeatother
\usepackage{xcolor}
\usepackage{longtable,booktabs,array}
\usepackage{calc} % for calculating minipage widths
\usepackage{etoolbox}
\makeatletter
\patchcmd\longtable{\par}{\if@noskipsec\mbox{}\fi\par}{}{}
\makeatother
\IfFileExists{footnotehyper.sty}{\usepackage{footnotehyper}}{\usepackage{footnote}}
\makesavenoteenv{longtable}
\usepackage{graphicx}
\makeatletter
\def\maxwidth{\ifdim\Gin@nat@width>\linewidth\linewidth\else\Gin@nat@width\fi}
\def\maxheight{\ifdim\Gin@nat@height>\textheight\textheight\else\Gin@nat@height\fi}
\makeatother
\setkeys{Gin}{width=\maxwidth,height=\maxheight,keepaspectratio}
\makeatletter
\def\fps@figure{htbp}
\makeatother
\providecommand{\tightlist}{%
  \setlength{\itemsep}{0pt}\setlength{\parskip}{0pt}}
\NewDocumentCommand\citeproctext{}{}
\NewDocumentCommand\citeproc{mm}{%
  \begingroup\def\citeproctext{#2}\cite{#1}\endgroup}
\makeatletter
 \let\@cite@ofmt\@firstofone
 \def\@biblabel#1{}
 \def\@cite#1#2{{#1\if@tempswa , #2\fi}}
\makeatother
\newlength{\cslhangindent}
\newlength{\csllabelwidth}
\newenvironment{CSLReferences}[2] % #1 hanging-indent, #2 entry-spacing
 {\begin{list}{}{%
  \setlength{\itemindent}{0pt}
  \setlength{\leftmargin}{0pt}
  \setlength{\parsep}{0pt}
  \ifodd #1
   \setlength{\leftmargin}{\cslhangindent}
   \setlength{\itemindent}{-1\cslhangindent}
  \fi
  \setlength{\itemsep}{#2\baselineskip}}}
 {\end{list}}
\usepackage{calc}

\ifLuaTeX
\usepackage[bidi=basic]{babel}
\else
\usepackage[bidi=default]{babel}
\fi
\babelprovide[main,import]{american}
\def\languageshorthands#1{}
\ifLuaTeX
  \usepackage{selnolig}  % disable illegal ligatures
\fi
\IfFileExists{bookmark.sty}{\usepackage{bookmark}}{\usepackage{hyperref}}
\IfFileExists{xurl.sty}{\usepackage{xurl}}{} % add URL line breaks if available
\hypersetup{
  pdftitle={cortecs: A Python package for compressing opacities},
  pdfauthor={Arjun B. Savel, Megan Bedell, Eliza M.-R. Kempton},
  pdflang={en-US},
  colorlinks=true,
  linkcolor={Maroon},
  filecolor={Maroon},
  citecolor={Blue},
  urlcolor={Blue},
  pdfcreator={LaTeX via pandoc}}

\title{cortecs: A Python package for compressing opacities}

\usepackage[affil-it]{authblk}
\usepackage{orcidlink}
\author[1,2%
  \ensuremath\mathparagraph]{Arjun B. Savel%
    \,\orcidlink{0000-0002-2454-768X}\,%
    }
\author[2%
  ]{Megan Bedell%
    \,\orcidlink{0000-0001-9907-7742}\,%
    }
\author[1%
  ]{Eliza M.-R. Kempton%
    \,\orcidlink{0000-0002-1337-9051}\,%
    }

\affil[1]{Astronomy Department, University of Maryland, College Park,
4296 Stadium Dr., College Park, MD 207842 USA}
\affil[2]{Flatiron Institute, Simons Foundation, 162 Fifth Avenue, New
York, NY 10010, USA}
\affil[$\mathparagraph$]{Corresponding author: %
}
\date{26 August 2023}

\begin{document}
\maketitle

\section{Summary}\label{summary}

The absorption and emission of light by exoplanet atmospheres encode
details of atmospheric composition, temperature, and dynamics.
Fundamentally, simulating these processes requires detailed knowledge of
the opacity of gases within an atmosphere. When modeling broad
wavelength ranges at high resolution, such opacity data, for even a
single gas, can take up multiple gigabytes of system random-access
memory (RAM). This aspect can be a limiting factor when considering the
number of gases to include in a simulation, the sampling strategy used
for inference, or even the architecture of the system used for
calculations. Here, we present \texttt{cortecs}, a Python tool for
compressing opacity data. \texttt{cortecs} provides flexible methods for
fitting the temperature, pressure, and wavelength dependencies of
opacity data and for evaluating the opacity with accelerated,
GPU-friendly methods. The package is actively developed on GitHub
(\url{https://github.com/arjunsavel/cortecs}), and it is available for
download with \texttt{pip} and \texttt{conda}.

\section{Statement of need}\label{statement-of-need}

Observations with the latest high-resolution spectrographs (e.g., IGRINS
/ Gemini South, ESPRESSO / VLT, MAROON-X / Gemini North; Mace et al.
(\citeproc{ref-mace:2018}{2018}); Seifahrt et al.
(\citeproc{ref-seifahrt:2020}{2020}); Pepe et al.
(\citeproc{ref-pepe:2021}{2021})) have motivated RAM-intensive
simulations of exoplanet atmospheres at high spectral resolution.
\texttt{cortecs} enables these simulations with more gases and on a
broader range of computing architectures by compressing opacity data.

Broadly, generating a spectrum to compare against recent high-resolution
data requires solving the radiative transfer equation over tens of
thousands of wavelength points (e.g., \citeproc{ref-beltz:2023}{Beltz et
al., 2023}; \citeproc{ref-gandhi:2023}{Gandhi et al., 2023};
\citeproc{ref-line:2021}{Line et al., 2021};
\citeproc{ref-maguire:2023}{Maguire et al., 2023};
\citeproc{ref-prinoth:2023}{Prinoth et al., 2023};
\citeproc{ref-savel:2022}{Savel et al., 2022};
\citeproc{ref-wardenier:2023}{Wardenier et al., 2023}). To decrease
computational runtime, some codes have parallelized the problem on GPUs
(e.g., \citeproc{ref-lee:2022}{Lee et al., 2022};
\citeproc{ref-line:2021}{Line et al., 2021}). However, GPUs cannot in
general hold large amounts of data in their video random-access memory
(VRAM) (e.g., \citeproc{ref-ito:2017}{Ito et al., 2017}); only the
cutting-edge, most expensive GPUs are equipped with VRAM in excess of 30
GB (such as the NVIDIA A100 or H100). RAM and VRAM management is
therefore a clear concern when producing high-resolution spectra.

How do we decrease the RAM footprint of these calculations? By far the
largest contributor to the RAM footprint, at least as measured on disk,
is the opacity data. For instance, the opacity data for a single gas
species across the wavelength range of the Immersion GRating INfrared
Spectrometer spectrograph (IGRINS, \citeproc{ref-mace:2018}{Mace et al.,
2018}) takes up 2.1 GB of non-volatile memory (i.e., the file size is
2.1 GB) at \texttt{float64} precision and at a resolving power of
400,000 (as used in Line et al. (\citeproc{ref-line:2021}{2021}); with
39 temperature points and 18 pressure points, using, e.g., the Polyansky
et al. (\citeproc{ref-polyansky:2018}{2018}) water opacity tables). In
many cases, not all wavelengths need to be loaded, e.g.~if the user is
down-sampling the resolution of their opacity function. Even so, it
stands to reason that decreasing the amount of RAM/VRAM consumed by
opacity data would strongly decrease the total amount of RAM/VRAM
consumed by the radiative transfer calculation.

One solution is to isolate redundancy: While the wavelength dependence
of opacity is sharp for many gases, the temperature and pressure
dependencies are generally smooth and similar across wavelengths (e.g.,
\citeproc{ref-barber:2014}{Barber et al., 2014};
\citeproc{ref-coles:2019}{Coles et al., 2019};
\citeproc{ref-polyansky:2018}{Polyansky et al., 2018}). This feature
implies that the opacity data should be compressible without significant
loss of accuracy at the spectrum level.

While our benchmark case (see the Benchmark section below) demonstrates
the applicability of \texttt{cortecs} to high-resolution opacity
functions of molecular gases, the package is general and the
compression/decompression steps of the package can be applied to any
opacity data in HDF5 format that has pressure and temperature
dependence, such as the opacity of neutral atoms or ions. Our benchmark
only shows, however, that the amounts of error from our compression
technique is reasonable in the spectra of exoplanet atmospheres at
pressures greater than a microbar for a single composition. This caveat
is important to note for a few reasons:

\begin{enumerate}
\def\labelenumi{\arabic{enumi}.}
\tightlist
\item
  Based on error propagation, the error in the opacity function may be
  magnified in the spectrum based on the number of cells that are traced
  during radiative transfer. The number of spatial cells used to
  simulate exoplanet atmospheres (in our case, 100) is small enough that
  the \texttt{cortecs} error is not large at the spectrum level.
\item
  Exoplanet atmospheres are often modeled in hydrostatic equilibrium at
  pressures greater than a microbar (e.g.,
  \citeproc{ref-barstow2020comparison}{Barstow et al., 2020};
  \citeproc{ref-showman2020atmospheric}{Showman et al., 2020}). When
  modeling atmospheres in hydrostatic equilibrium, the final spectrum
  essentially maps to the altitude at which the gas becomes optically
  thick. If \texttt{cortecs}-compressed opacities were used to model an
  optically thin gas over large path lengths, however, then smaller
  opacities would be more important. \texttt{cortecs} tends to perform
  worse at modeling opacity functions that jump from very low to very
  high opacities, so it may not perform optimally in these optically
  thin scenarios.
\item
  The program may perform poorly for opacity functions with sharp
  features in their temperature--pressure dependence (e.g., the Lyman
  series transitions of hydrogen,
  \citeproc{ref-kurucz2017including}{Kurucz, 2017}). That is, the data
  may require so many parameters to be fit that the compression is no
  longer worthwhile.
\end{enumerate}

\section{Methods}\label{methods}

\texttt{cortecs} seeks to compress redundant information by representing
opacity data not as the opacity itself but as fits to the opacity as a
function of temperature and pressure. We generally refer to this process
as \emph{compression} as opposed to \emph{fitting} to emphasize that we
do not seek to construct physically motivated, predictive, or
comprehensible models of the opacity function. Rather, we simply seek
representations of the opacity function that consume less RAM/VRAM. The
compression methods we use are \emph{lossy} --- the original opacity
data cannot be exactly recovered with our methods. We find that the loss
of accuracy is tolerable for at least the hot Jupiter emission
spectroscopy application (see Benchmark below).

\begin{figure}
\centering
\includegraphics{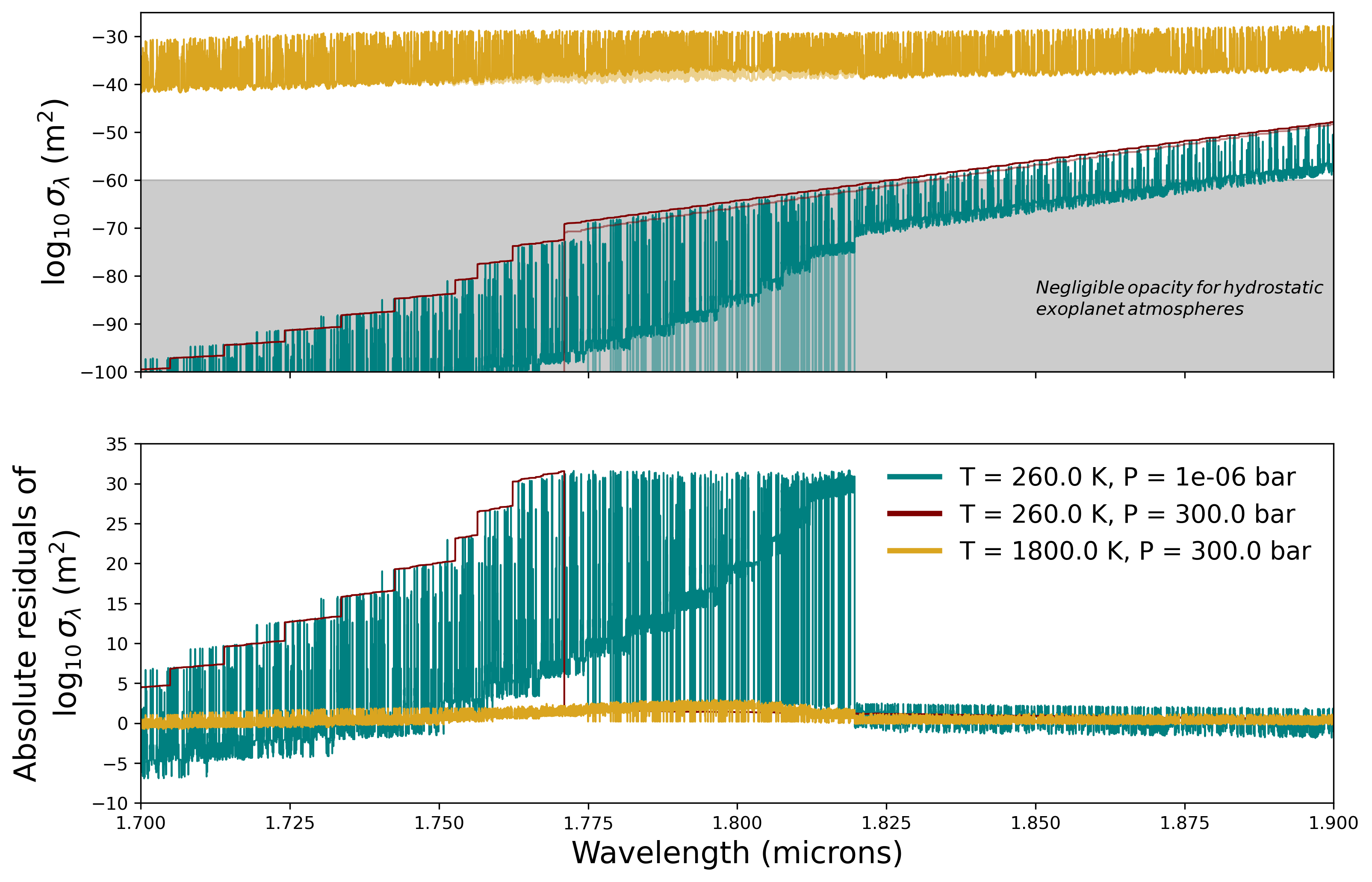}
\caption{Top panel: The original opacity function of CO
(\citeproc{ref-rothman:2010}{Rothman et al., 2010}) (solid lines) and
its \texttt{cortecs} reconstruction (transparent lines) over a large
wavelength range and at multiple temperatures and pressures. Bottom
panel: the absolute residuals between the opacity function and its
\texttt{cortecs} reconstruction. \(\sigma_\lambda\) is the opacity, in
units of square meters. We cut off the opacity at \(10^{-104}\),
explaining the shape of the residuals in teal and dark red. Note that
opacities less than \(10^{-60}\) are not generally relevant for the
benchmark presented here; an opacity of \(\sigma_\lambda=10^{-60}\)
would require a column nearly \(10^{34}\)m long to become optically
thick at a pressure of 1 bar and temperature of 1000 K.
\label{fig:example}}
\end{figure}

We provide three methods of increasing complexity (and flexibility) for
compressing and decompressing opacity: polynomial-fitting, principal
components analysis (PCA, e.g., \citeproc{ref-jolliffe:2016}{Jolliffe \&
Cadima, 2016}) and neural networks (e.g.,
\citeproc{ref-alzubaidi:2021}{Alzubaidi et al., 2021}). The default
neural network architecture is a fully connected neural network; the
user can specify the desired hyperparameters, such as number of layers,
neurons per layer, and activation function. Alternatively, any
\texttt{keras} model (\citeproc{ref-chollet:2015}{Chollet, 2015}) can be
passed to the fitter. Each compression method is paired with a
decompression method for evaluating opacity as a function of
temperature, pressure, and wavelength. These decompression methods are
tailored for GPUs and are accelerated with the \texttt{JAX} code
transformation framework (\citeproc{ref-jax:2018}{Bradbury et al.,
2018}). An example of this reconstruction is shown in
\autoref{fig:example}. In the figure, opacities less than \(10^{-60}\)
are ignored. This is because, to become optically thick at a pressure of
1 bar and temperature of 1000 K, a column would need to be nearly
\(10^{34}\)m long. Here we show a brief derivation of this. The length
of the column, \(ds\) is \(ds = \frac{\tau}{\alpha}\), where \(\tau\) is
the optical depth, and \(\alpha\) is the absorption coefficient. Setting
\(\tau = 1\), we have \(ds = \frac{1}{\alpha}\). The absorption
coefficient is the product of the opacity and the density of the gas:
\(ds = \frac{1}{\kappa_\lambda \rho}\).
Therefore,\(ds = \frac{1}{\kappa_\lambda \rho}\). The density of the gas
\(\rho\) is the pressure divided by the product of the temperature and
the gas constant: \(\rho = \frac{P}{k_B T \mu}\) for mean molecular
weight \(\mu\). This leads to the final equation for the column length:
\(ds = \frac{k_BT\mu}{P\kappa_\lambda}\). For CO, the mean molecular
weight is 28.01 g/mol. Plugging in, we arrive at \(ds \approx 10^{34}\)m
(roughly \(10^{17}\) parsecs) for \(\kappa_\lambda = 10^{-33}\)
\(\rm cm^2/g\), which is equivalent to roughly a cross-section of
\(\sigma_\lambda = 10^{-60}\) \(\rm m^2\).

\section{Workflow}\label{workflow}

A typical workflow with \texttt{cortecs} involves the following steps:

\begin{enumerate}
\def\labelenumi{\arabic{enumi}.}
\tightlist
\item
  Acquiring: Download opacity data from a source such as the ExoMol
  database (\citeproc{ref-tennyson:2016}{Tennyson et al., 2016}) or the
  HITRAN database (\citeproc{ref-gordon:2017}{Gordon et al., 2017}).
\item
  Fitting: Compress the opacity data with \texttt{cortecs}'s
  \texttt{fit} methods.
\item
  Saving: Save the compressed opacity data (the fitted parameters) to
  disk.
\item
  Loading: Load the compressed opacity data from disk in whatever
  program you're applying the data---e.g., within your radiative
  transfer code.
\item
  Decompressing: Evaluate the opacity with \texttt{cortecs}'s
  \texttt{eval} methods.
\end{enumerate}

The accuracy of these fits may or may not be suitable for a given
application. It is important to test that the error incurred using
\texttt{cortecs} does not impact the results of your application---for
instance, by using the \texttt{cortecs.fit.metrics.calc\_metrics}
function to calculate the error incurred by the compression and by
calculating spectra with and without using \texttt{cortecs}-compressed
opacities. We provide an example of such a benchmarking exercise below.

\section{Benchmark: High-resolution retrieval of
WASP-77Ab}\label{benchmark-high-resolution-retrieval-of-wasp-77ab}

As a proof of concept, we perform a parameter inference exercise (a
``retrieval,'' \citeproc{ref-madhusudhan:2009}{Madhusudhan \& Seager,
2009}) on the high-resolution thermal emission spectrum of the fiducial
hot Jupiter WASP-77Ab (\citeproc{ref-august:2023}{August et al., 2023};
\citeproc{ref-line:2021}{Line et al., 2021};
\citeproc{ref-mansfield:2022}{Mansfield et al., 2022}) as observed at
IGRINS. The retrieval pairs \texttt{PyMultiNest}
(\citeproc{ref-buchner:2014}{Buchner et al., 2014}) sampling with the
\texttt{CHIMERA} radiative transfer code (\citeproc{ref-line:2013}{Line
et al., 2013}), with opacity from \(\rm H_2O\)
(\citeproc{ref-polyansky:2018}{Polyansky et al., 2018}), \(\rm CO\)
(\citeproc{ref-rothman:2010}{Rothman et al., 2010}), \(\rm CH_4\)
(\citeproc{ref-hargreaves:2020}{Hargreaves et al., 2020}), \(\rm NH_3\)
(\citeproc{ref-coles:2019}{Coles et al., 2019}), \(\rm HCN\)
(\citeproc{ref-barber:2014}{Barber et al., 2014}), \(\rm H_2S\)
(\citeproc{ref-azzam:2016}{Azzam et al., 2016}), and \(\rm H_2-H_2\)
collision-induced absorption (\citeproc{ref-karman:2019}{Karman et al.,
2019}). The non-compressed retrieval uses the data and retrieval
framework from (\citeproc{ref-line:2021}{Line et al., 2021}), run in an
upcoming article (Savel et al.~2024, submitted). For this experiment, we
use the PCA-based compression scheme implemented in \texttt{cortecs},
preserving 2 principal components and their corresponding weights as a
function for each wavelength as a lossy compression of the original
opacity data.

Using \texttt{cortecs}, we compress the input opacity files by a factor
of 13. These opacity data (as described earlier in the paper) were
originally stored as 2.1 GB .h5 files containing 39 temperature points,
18 pressure points, and 373,260 wavelength points. The compressed
opacity data are stored as a 143.1 MB .npz file, including the PCA
coefficients and PCA vectors (which are reused for each wavelength
point). These on-disk memory quotes are relatively faithful to the
in-memory RAM footprint of the data when stored as \texttt{numpy} arrays
(2.1 GB for the uncompressed data and 160 MB for the compressed data).
Reading in the original files takes 1.1 \(\pm\) 0.1 seconds, while
reading in the compressed files takes 24.4 \(\pm\) 0.3 ms.
Accessing/evaluating a single opacity value takes 174.0 \(\pm\) 0.5 ns
for the uncompressed data and 789 \(\pm\) 5 ns for the compressed data.
All of these timing experiments are performed on a 2021 MacBook Pro with
an Apple M1 Pro chip and 16 GB of RAM.

\begin{figure}
\centering
\includegraphics{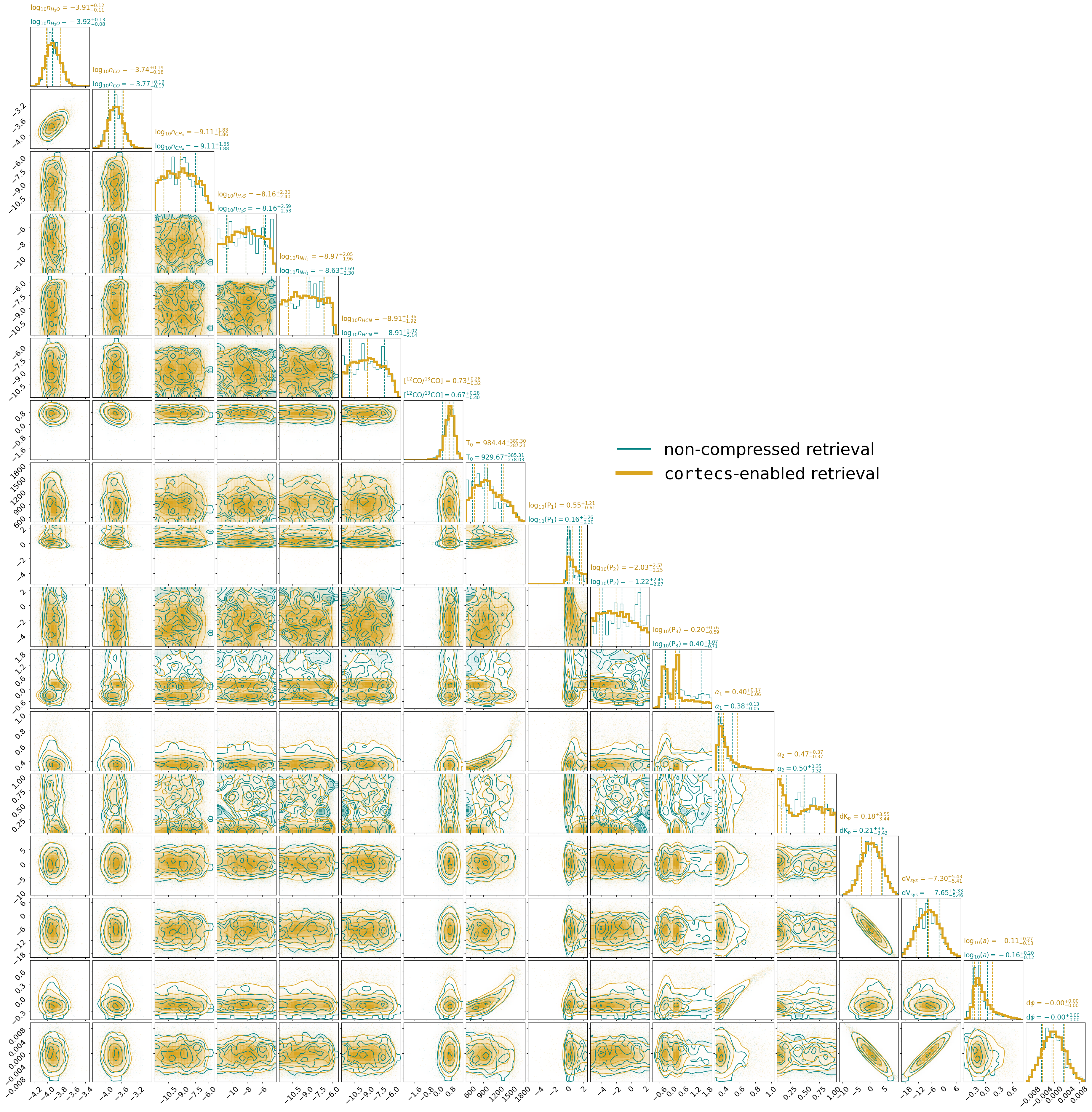}
\caption{The posterior distributions for our baseline WASP-77Ab
retrieval (teal) and our retrieval using opacities compressed by
\texttt{cortecs} (gold). \label{fig:corner}}
\end{figure}

Importantly, we find that our compressed-opacity retrieval yields
posterior distributions (as plotted by the \texttt{corner} package,
\citeproc{ref-corner:2016}{Foreman-Mackey, 2016}) and Bayesian evidences
that are consistent with those from the benchmark retrieval using
uncompressed opacity (\autoref{fig:corner}) within a comparable runtime.
The two posterior distributions exhibit slightly different substructure,
which we attribute to the compressed results requiring 10\% more samples
to converge (about 5 hours of extra runtime on a roughly 2 day-long
calculation) and residual differences between the compressed and
uncompressed opacities. The results from this exercise indicate that our
compression/decompression scheme is accurate enough to be used in at
least some high-resolution retrievals.

\begin{longtable}[]{@{}
  >{\raggedright\arraybackslash}p{(\columnwidth - 8\tabcolsep) * \real{0.1468}}
  >{\raggedright\arraybackslash}p{(\columnwidth - 8\tabcolsep) * \real{0.1835}}
  >{\raggedright\arraybackslash}p{(\columnwidth - 8\tabcolsep) * \real{0.2477}}
  >{\raggedright\arraybackslash}p{(\columnwidth - 8\tabcolsep) * \real{0.2018}}
  >{\raggedright\arraybackslash}p{(\columnwidth - 8\tabcolsep) * \real{0.2202}}@{}}
\toprule\noalign{}
\begin{minipage}[b]{\linewidth}\raggedright
Method
\end{minipage} & \begin{minipage}[b]{\linewidth}\raggedright
Compression factor
\end{minipage} & \begin{minipage}[b]{\linewidth}\raggedright
Median absolute deviation
\end{minipage} & \begin{minipage}[b]{\linewidth}\raggedright
Compression time (s)
\end{minipage} & \begin{minipage}[b]{\linewidth}\raggedright
Decompression time (s)
\end{minipage} \\
\midrule\noalign{}
\endhead
\bottomrule\noalign{}
\endlastfoot
PCA & 13 & 0.30 & 2.6 \(\times 10^1\) & 2.3 \(\times 10^2\) \\
Polynomials & 44 & 0.24 & 7.8\(\times 10^2\) & 3.6\(\times 10^3\) \\
Neural network & 9 & 2.6 & 1.4\(\times 10^7\) & 3.6\(\times 10^4\) \\
\end{longtable}

Comparison of compression methods used for the full HITEMP CO line list
(\citeproc{ref-rothman:2010}{Rothman et al., 2010}) over the IGRINS
wavelength range at a resolving power of 250,000, cumulative for all
data points. Note that the neural network compression performance and
timings are only assessed at a single wavelength point and extrapolated
over the full wavelength range.

\section{Acknowledgements}\label{acknowledgements}

A.B.S. and E.M-R.K. acknowledge support from the Heising-Simons
Foundation. We thank Max Isi for helpful discussions.

\section*{References}\label{references}
\addcontentsline{toc}{section}{References}

\phantomsection\label{refs}

\end{document}